\documentclass[journal]{IEEEtran}
\usepackage[utf8]{inputenc}
\usepackage{stfloats}
\usepackage{cite}
\usepackage{amsmath}
\usepackage{amsfonts}
\usepackage{dsfont}
\usepackage{graphicx}
\usepackage{changepage}
\usepackage{color}
\usepackage{epstopdf}
\UseRawInputEncoding

\ifCLASSINFOpdf

\fi

\begin{document}

\title{On the Performance of RIS-Assisted Dual-Hop Mixed RF-UWOC Systems}

\author{Sai Li, Liang Yang, Daniel Benevides da Costa, Marco Di Renzo, and Mohamed-Slim Alouini
\thanks{S. Li and L. Yang are with the College of Computer Science and Electronic Engineering, Hunan University, Changsha 410082,
China, (e-mail:lisa2019@hnu.edu.cn, liangy@hnu.edu.cn).}
\thanks{D. B. da Costa is with the Department of Computer Engineering, Federal University of Cear\'{a}, Sobral 62010-560, Brazil (email: danielbcosta@ieee.org).}
\thanks{M. Di Renzo is with Universit\'{e} Paris-Saclay, CNRS and CentraleSup\'{e}lec, Laboratoire des Signaux et Syst\`{e}mes, Gif-sur-Yvette, France. (email: marco.direnzo@centralesupelec.fr).}
\thanks{M.-S. Alouini is with the CEMSE Division, King Abdullah University of Science and Technology (KAUST), Thuwal 23955-6900, Saudi Arabia (email: slim.alouini@kaust.edu.sa).}}
\maketitle

\begin{abstract}
In this paper, we investigate the performance of a reconfigurable intelligent surface (RIS)-assisted dual-hop mixed radio-frequency underwater wireless optical communication (RF-UWOC) system. An RIS is an emerging and low-cost technology that aims to enhance the strength of the received signal, thus improving the system performance. In the considered system setup, a ground source does not have a reliable direct link to a given marine buoy and communicates with it through an RIS installed on a building. In particular, the buoy acts as a relay that sends the signal to an underwater
destination. In this context, analytical expressions for the outage probability (OP), average bit error rate (ABER), and average channel capacity (ACC) are derived assuming fixed-gain amplify-and-forward (AF) and decode-and-forward (DF) relaying protocols at the marine buoy.
Moreover, asymptotic analyses of the OP and ABER are carried out in order to gain further insights from the analytical frameworks. In particular, the system diversity order is derived and it is shown to depend on the RF link parameters and on the detection schemes of the UWOC link. Finally, it is demonstrated that RIS-assisted systems can effectively improve the performance of mixed dual-hop RF-UWOC systems.
\end{abstract}

\begin{IEEEkeywords}
Underwater wireless optical communications, reconfigurable intelligent surface, relaying systems, mixed dual-hop transmission schemes.
\end{IEEEkeywords}

\section{Introduction}
\IEEEPARstart{U}{nderwater} wireless optical communication (UWOC) is a promising technique for beyond fifth-generation (B5G) wireless networks because it provides higher data rates and better confidentiality than traditional underwater wireless communication systems \cite{1,2}.
In UWOC, however, the scattering and absorption of the optical signal caused by various components in the seawater may have an adverse effect on the transmission of the optical signal, and these issues need to be properly evaluated. Along the years, mixed dual-hop transmissions assuming radio-frequency (RF) and UWOC links have been widely investigated. Specifically, the secrecy performance of mixed RF-UWOC systems was investigated in \cite{3}. In \cite{4}, mixed RF-UWOC relaying networks with both decode-and-forward (DF) and amplify-and-forward (AF) capabilities were analyzed. In \cite{5}, the authors proposed a unified UWOC channel model that accounts for the fluctuations of the beam irradiance in the presence of both air bubbles and temperature gradients. Such a distribution was called the Exponential-Generalized Gamma (EGG) distribution and it encompasses as a special case the Exponential Gamma (EG) fading model at a uniform temperature. Furthermore, in \cite{5}, closed-form expressions for the outage probability (OP), average bit error rate (ABER), and average channel capacity (ACC) were derived. Using the EGG distribution, the performance of dual-hop UWOC systems with a fixed-gain AF relay was studied in \cite{6}. Finally, relying on the EGG fading model, the performance of mixed RF-UWOC systems was investigated in \cite{7,8,9}.

Reconfigurable intelligent surface (RIS) technology is a promising efficient solution to overcome the negative effects of wireless channels, by making the wireless environment programmable. By smarting controlling the polarization, scattering, reflection, and refraction of the impinging
electromagnetic waves, the propagation environment can be customized in order to meet the desired application requirements \cite{10,11,12,13,14,15,16}. In \cite{16}, a detailed overview of the modeling and application of RISs in wireless communications was provided, in which the differences between RISs and other related techniques, such as relays, were discussed and the Non-Central Chi-Square (NCCS) distribution was employed for performance analysis. Channel modeling and beamforming optimization design issues of RIS-assisted systems were studied in \cite{17,18,19,20,21}. The performance of RIS-assisted mixed free-space optical-RF (FSO-RF) systems was examined in \cite{22}, while in \cite{23} the authors studied the coverage and signal-to-noise ratio (SNR) gain of RIS-assisted communication systems. In addition, the secrecy performance of RIS-assisted secure wireless communication systems was studied in \cite{24,25}.

Although the NCCS distribution has recently emerged as an efficient statistical model for RIS-assisted systems, such a model has some limitations. In the high SNR regime, for example, the deviation between simulation and theoretical
results is large. In addition, the NCCS model is only applicable to scenarios with a large number of reflecting elements $N$. In order to overcome the limitations of using the NCCS distribution, a more accurate distribution was proposed in \cite{26} to approximate the end-to-end channel distribution of RIS-aided wireless systems, which is referred to as the squared generalized-K ($K_G$) distribution. By relying upon the $K_G$ distribution, analytical and simulation results well match for an arbitrary number of reflecting elements $N$. The approach proposed in \cite{23} was applied to the performance evaluation of unmanned aerial vehicle (UAV) relaying systems \cite{27} and indoor mixed visible light communication/RF (VLC/RF) systems \cite{28}.

As discussed in previous text, RISs have been leveraged for enhancing the performance of various emerging technologies, such as FSO, UAV, and VLC networks \cite{21,22,27,28}. Moreover, RISs have bee applied for improving the performance of multiple-input multiple/single-output (MIMO/MISO) networks \cite{20,29,30} and millimeter wave (mmWave) communications \cite{31,32}. Furthermore, dual-hop mixed RF-UWOC systems have been employed in applications like ocean monitoring \cite{4}. However, the application of RISs in dual-hop mixed RF-UWOC communication systems has not been studied yet. Motivated by these considerations, in this paper we investigate the scenario in which an RIS is used to reflect the signals transmitted from a terrestrial source to a buoy relay located on the sea surface, which in turn relays the signal to an underwater destination by using fixed gain AF or DF relaying protocols. To evaluate the impact of key system parameters on the overall performance, which include the number of reflecting elements, as well as the bubble level, temperature gradient, water type, and detection technology, a comprehensive performance analysis is carried out in terms of OP, ABER, and ACC.
The main contributions of this paper can be summarized as follows:

$\bullet$ We study an RIS-aided dual-hop mixed RF-UWOC system, where the RF and UWOC links are modeled by using a squared $K_G$ distribution and an EGG distribution, respectively. Both heterodyne detection (HD) and intensity modulation/direct detection (IM/DD) techniques are considered for the optical link.

$\bullet$ Closed-form expressions for the cumulative distribution function (CDF) and probability density function (PDF) of the end-to-end (e2e) SNR under both fixed-gain AF and DF relaying protocols are derived. Based on these distributions, a performance analysis is carried out and closed-form expressions are derived for the OP, ABER, and ACC.

$\bullet$ To get further insights from the analytical frameworks, asymptotic expressions for the OP and ABER are derived in the form of simple elementary functions, which enables us to determine the system diversity order. It is shown that the achievable diversity order depends on the RF link parameters as well as on the type of the detection technique at the UWOC link.

$\bullet$ Insightful discussions are provided. For instance, it is found that when the bubble level and temperature gradient are low, that is, when the underwater turbulence intensity is small, the overall system performance is better. In addition, it is shown that the e2e performance of the UWOC link is better if the HD scheme is used. Finally, it is demonstrated through several illustrative numerical examples that RIS-assisted schemes can effectively improve the performance of mixed dual-hop RF-UWOC systems.

The remainder of this paper is organized as follows. Section II introduces the system and channel models along with some useful statistics for the RF and UWOC links. In Section III, a more in-depth statistical characterization is performed for both fixed-gain AF and DF relaying schemes.
Based on the preliminary results obtained in the previous sections, Section IV derives exact closed-form expressions for key performance metrics and an asymptotic analysis is conducted as well. Section VI provides some illustrative numerical examples based on which insightful findings are obtained. Finally, Section VII concludes the paper.

\begin{figure}[t]
    \centering
    \includegraphics[width=3.5in]{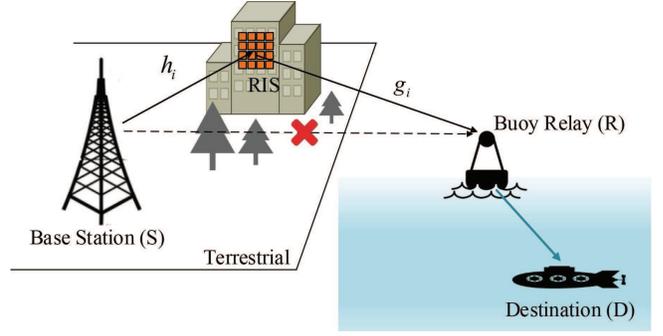}
    \caption{The RIS-assisted dual-hop mixed RF-UWOC system model.}
\end{figure}

\section{System and Channel Models}
We consider an RIS-assisted dual-hop mixed RF-UWOC system that consists of a base station (S), an RIS with $N$ reflecting elements that is
installed on a building, a buoy relay (R) that is located on the water surface, and an underwater destination (D), as shown in Fig. 1. It is assumed that R can operate under both fixed-gain AF and DF relaying protocols. All nodes are equipped with a single antenna, and there is no direct link between S and R due to the long distance and the presence of environmental obstructions. Therefore, the RF signal is first sent from S to the RIS, and then the RIS reflects the incoming signal to R. After the electro-optical conversion at R, the optical signal is transmitted to D via the UWOC channel. In Fig. 1, $h_{i}=\alpha_{i}e^{-j\theta_{i}}$ and $g_{i}=\beta_{i}e^{-j\varphi_{i}}$ are, respectively, the fading channels of the S-RIS and RIS-R links related to the $i$th reflecting element ($i=1,2,\ldots,N$), where $\alpha_{i}$ and $\theta_{i}$ denote the channel amplitude and phase of $h_{i}$, respectively, while $\beta_{i}$ and $\varphi_{i}$ represent the channel amplitude and phase of $g_{i}$, respectively.
The RIS provides adjustable phase shifts and, in our analysis, the channel phases of $h_{i}$ and $g_{i}$ at the RIS are assumed to be known, which corresponds to the best scenario in terms of system operation. In addition, from a practical point of view, due to the high probability of line-of-sight signals on the S-RIS link, it is assumed that the S-RIS link obeys a Nakagami-$m$ fading distribution. Furthermore, since the RIS-R link is usually impaired by complex near-ground environmental factors (e.g., the presence of trees or buildings), we assume that the RIS-R link follows a Rayleigh fading.

\subsection{The SNR of the RF Link}
From [22, Eq. (1)],  the instantaneous SNR at R, $\gamma_{1}$, can be formulated as follows
\begin{align}
\gamma_{1}=\Big|\sum_{i=1}^{N}h_{i}v_{i}g_{i}\Big|^{2}\overline\gamma_{1},\tag{1}\label{1}
\end{align}
where $\overline\gamma_{1}$ is the average SNR of the RF link,
$v_{i}=\rho_{i}(\phi_{i})e^{\phi_{i}}$ is the reflection coefficient introduced by the $i$th reconfigurable element of the RIS, and $\rho_{i}(\phi_{i})=1$, $\forall i$, for ideal reflecting elements. When $\rho_{i}(\phi_{i})=1$, (\ref{1}) reduces to $\gamma_{1}=|\sum_{i=1}^{N}\alpha_{i}\beta_{i}e^{(\phi_{i}-\theta_{i}-\varphi_{i})}|^{2}\overline\gamma_{1}$.
In order to obtain the maximum value of $\gamma_{1}$, the reflected signals need to be co-phased by setting $\phi_{i}=\theta_{i}+\varphi_{i}$, for $i=1,2,\ldots,N$.

Therefore, the maximum instantaneous SNR at R can be written as \cite{16,26}
\begin{align}
\gamma_{1}=Z^2\overline\gamma_{1},\tag{2}\label{2}
\end{align}
where $Z=\sum_{i=1}^{N}\alpha_{i}\beta_{i}$, $\alpha_{i}$ is a Nakagami-$m$ random variable (RV), with $m\geq\frac{1}{2}$ being the fading parameter of the distribution, and $\beta_{i}$ is a Rayleigh RV with mean $\sqrt{\pi}/2$ and variance $(4-\pi)/4$. By letting $\chi_{i}=\alpha_{i}\beta_{i}$, $Z$ turns out to be the sum of $N$ independent and identically distributed (i.i.d.) RVs $\chi_{i}$. The PDF of $\chi_{i}$ can be proved to be $f_{\chi_{i}}(x)=\frac{4m^\frac{1+m}{2}x^m}
{\Gamma(m)}K_{m-1}(2\sqrt{m}x)$ [27, Eq. (5)], [33, Eq. (8.432.7)], where $K_{v}(\cdot)$ is the modified Bessel function [33, Eq. (8.432)] and $\Gamma(\cdot)$ denotes the Gamma function [33, Eq. (8.310)]. Comparing the PDF of $\chi_{i}$ with [34, Eq. (1)], one can see that the PDF of $\chi_{i}$ is a special case of the $K_G$ distribution. In \cite{34}, it is
stated that the PDF of the sum of multiple $K_{G}$ RVs can be well-approximated by the PDF of $\sqrt{W}$, with $W = \sum_{j=1}^{N}\chi_{i}^2$. Therefore, the PDF of $Z$ can be formulated as
\begin{equation}
f_{Z}(x) = \frac{4\widetilde{\Xi}^{\frac{k_{w}+m_{w}}{2}}}
{\Gamma (k_{w})\Gamma (m_{w})}x^{k_{w}+m_{w}-1}K_{k_{w}-m_{w}}(2\sqrt{\widetilde{\Xi}} x),\tag{3}\label{3}
\end{equation}
where $k_{w}=\frac{-b_{w}+\sqrt{b_{w}^2-4a_{w}c_{w}}}{2a_{w}}$ and $m_{w}=\frac{-b_{w}-\sqrt
{b_{w}^2-4a_{w}c_{w}}}{2a_{w}}$ are the the shaping parameters, $\widetilde{\Xi} = k_{w}m_{w}
/\Omega _{w}$, and $\Omega _{w} = \mu _{R}(2)$
is the mean power of $Z$. Moreover, the parameters $a_{w}$, $b_{w}$, and $c_{w}$ are
defined in \cite{34} and they are related to the moment $\mu _{Z}(n)$ of $Z$, namely,
\begin{align}
\mu _{Z}(n)&= \sum_{n_{1}=0}^{n}\sum_{n_{2}=0}^{n_{1}}...\sum_{n_{N-1}=0}^
{n_{N-2}}\binom{n}{n_{1}}\binom{n_{1}}{n_{2}}...\binom{n_{N-2}}{n_{N-1}} \nonumber\\
 &\times \mu _{\chi_{1}}(n-n_{1})\mu _{\chi_{2}}(n_{1}-n_{2})...\mu _{\chi_{N-1}}(n_{N-1}),\tag{4}\label{4}
\end{align}
where $\mu _{\chi_{i}}(n){=}\frac{2^{m{+}n{+}1}m^{\frac{1{+}m}{2}}}{\Gamma(m)}
\Gamma\left(\frac{n}{2}\right)\Gamma\left(m{+}\frac{n}{2}\right)$ is the $n$th
moment of $\chi_{i}$. Note that $k_{w}$ and $m_{w}$ are required to be real numbers.
But, in some cases, when $k_{w}$ and $m_{w}$ are calculated by using the moment-based estimators,
they are conjugate complex numbers. In this case, they
are set to the estimated modulus values of the conjugate complex number.

From (\ref{2}) and (\ref{3}), the PDF of the instantaneous SNR, $\gamma_{1}$, is given by
\begin{align}
f_{\gamma_{1}}(\gamma_{1})=\frac{2\Xi^{\frac{k_{w}{+}m_{w}}{2}}\gamma_{1}
^{\left(\frac{k_w{+}m_w}{2}{-}1\right)}}{\Gamma(k_w)\Gamma(m_w)}K_{k_w{-}m_w}
\left(2\sqrt{\Xi\gamma_{1}}\right),\tag{5}\label{5}
\end{align}
where $\Xi=\widetilde{\Xi}/\overline\gamma_{1}$. With the aid of [35, Eq. (8.4.23.1)] and [33, Eq. (9.31.5)],
and after some algebraic manipulations, the PDF in (5) can be rewritten as
\begin{align}
f_{\gamma_{1}}(\gamma_{1})=\frac{\widetilde{\Xi}}{\overline\gamma_{1}\Gamma(m_w)\Gamma(k_w)}
\mathrm{G}_{0,2}^{2,0}\left [\left.\begin{matrix}\frac{\widetilde{\Xi}\gamma_{1}}{\overline\gamma_{1}}
\end{matrix}\right|\begin{matrix}-\\ k_w{-}1,m_w{-}1 \end{matrix}\right],\tag{6}\label{6}
\end{align}
where $\mathrm{G}_{\cdot,\cdot}^{\cdot,\cdot}[\cdot]$ denotes the Meijer's G-function [33, Eq. (9.301)].
Thus, the CDF of $\gamma_{1}$ can be formulated as
\begin{align}
F_{\gamma_{1}}(\gamma_{1})=\frac{1}{\Gamma(m_w)\Gamma(k_w)}
\,{\mathrm{G}}_{1,3}^{2,1}\left [\left.\begin{matrix}\frac{\widetilde{\Xi}\gamma_{1}}{\overline\gamma_{1}}
\end{matrix}\right|\begin{matrix}1\\ k_w,m_w,0 \end{matrix}\right]\!.\tag{7}\label{7}
\end{align}
According to (4), we find that $m_w$ and $k_w$ depend on $N$ and $m$, with $k_w>m_w$.
After a careful inspection, when $m\geq0.5$ and $m_w\geq1$, it follows that $m_w$ increases as $N$ and $m$ increase.
Therefore, the CDF and PDF of $\gamma_{1}$ are directly related to $N$ and $m$.

\subsection{The SNR of the UWOC Link}
The UWOC link is modelled as a mixture EGG distribution which is equivalent to a weighted sum of the Exponential and Generalized Gamma (GG)
distributions \cite{5}. The instantaneous SNR of the UWOC link, $\gamma_{2}$, is given by
\begin{align}
\gamma_{2}=\mu_{r}I^r,\tag{8}\label{8}
\end{align}
where $r$ is the parameter that depends on the detection technique used, i.e., $r=1$ if the HD technique is considered and $r=2$
if the IM/DD technique is considered, $I$ denotes the channel coefficient that is distributed according to a mixture EGG distribution, and $\mu_{r}$ is the average electrical SNR. As for the HD technique, $\mu_{1}=\overline\gamma_{2}$. As for the IM/DD technique, $\mu_{2}=\frac{\overline\gamma_{2}}{2\omega
\lambda^{2}+b^{2}(1-\omega)\Gamma(a+2/c)/\Gamma(a)}$ [5, Eq. (19)], where $\overline\gamma_{2}$ represents the average SNR of the
UWOC link, $0<\omega<1$ is the mixture weight or mixture coefficient of the distributions, $\lambda$ is the fading
parameter of the Exponential distribution, and $a$, $b$, and $c$ represent the fading parameters associated with the GG distribution.
These parameters vary with the water temperature, water salinity, and bubble level. Typical parameters of the EGG distribution are
given in Table I and Table II.

From [5, Eqs. (21) and (22)], the PDF and CDF of $\gamma_{2}$ can be written in terms of Meijer's G-function. Departing from this representation and relying on [36, Eq. (07.34.26.0008.01)] and [37, Eq. (1.59)], the unified PDF and CDF expressions of $\gamma_{2}$ can be formulated in terms of Fox's H-function as
\begin{align}
f_{\gamma_{2}}(\gamma_{2})=\frac{\omega}{\gamma_{2}}&\, {\mathrm{H}}_{0,1}^{1,0}
\left [{{\frac{\gamma_{2}}{\lambda^r \mu_{r}}}\left |{ \begin{matrix} {-}
\\ {(1,r)} \\ \end{matrix} }\right . }\right ]\!\nonumber\\
&+\frac{(1-\omega)}{\gamma_{2}\Gamma(a)}\, {\mathrm{H}}_{0,1}^{1,0}
\left [{{\frac{\gamma_{2}}{b^r \mu_{r}}}\left |{ \begin{matrix} {-}
\\ {(a,\frac{r}{c})} \\ \end{matrix} }\right . }\right ]\!,\tag{9}\label{9}
\end{align}
\begin{align}
F_{\gamma_{2}}(\gamma_{2})=\omega r&\, {\mathrm{H}}_{1,2}^{1,1}
\left [{{\frac{\gamma_{2}}{\lambda^r \mu_{r}}}\left |{ \begin{matrix} {(1,r)}
\\ {(1,r)(0,r)} \\ \end{matrix} }\right . }\right ]\!\nonumber\\
&+\frac{(1-\omega)r}{\Gamma(a)c}\, {\mathrm{H}}_{1,2}^{1,1}
\left [{{\frac{\gamma_{2}}{b^r \mu_{r}}}\left |{ \begin{matrix} {(1,\frac{r}{c})}
\\ {(a,\frac{r}{c})(0,\frac{r}{c})} \\ \end{matrix} }\right . }\right ]\!,\tag{10}\label{10}
\end{align}
where $\mathrm{H}_{\cdot,\cdot}^{\cdot,\cdot}[\cdot]$ is the Fox's H-function [37, Eq. (1.2)].

\begin{table}[t]
\centering
\caption{Parameters of the EGG Distribution for Different Bubble Levels BL (L/min)
and Temperature Gradient $\Delta \mathrm{T}$ ($^{\circ}\mathrm{C}.cm^{-1}$) \cite{5}}
{\begin{tabular}{p{0.8cm}p{1.3cm}p{0.6cm}p{0.6cm}p{0.6cm}p{0.6cm}p{0.7cm}}
 \hline
 \hline
BL (L/min) & $\Delta \mathrm{T}$ $ (^{\circ}\mathrm{C}.cm^{-1})$  & $\omega$ &$\lambda$ & a & b & c\\
 \hline
2.4 & 0.05 & 0.2130 & 0.3291 & 1.4299 & 1.1817 & 17.1984 \\
2.4 & 0.20 & 0.1665 & 0.1207 & 0.1559 & 1.5216 & 22.8754 \\
4.7 & 0.05 & 0.4580 & 0.3449 & 1.0421 & 1.5768 & 35.9424 \\
 \hline
 \hline
\end{tabular}}
\end{table}

\section{Closed-Form End-to-End Statistics}
In this section, we derive closed-form expressions for the e2e statistics (i.e., CDF and PDF of the SNR)
of RIS-assisted dual-hop mixed RF-UWOC system assuming both fixed-gain AF and DF relaying protocols.
These expressions are useful for the performance analysis carried out in the next section.
\subsection{Fixed-Gain AF Relaying Protocol}
As for a fixed-gain AF relaying protocol, the overall e2e instantaneous SNR at D can be written as \cite{38}
\begin{align}
\gamma^{AF}=\frac{\gamma_{1}\gamma_{2}}{\gamma_{2}+C},\tag{11}\label{11}
\end{align}
where $C$ is a constant related to the relay gain.
\subsubsection{Cumulative Distribution Function}
The CDF of the e2e SNR can be expressed as
\begin{align}
F_{\gamma^{AF}}(\gamma)&=\int_{0}^{\infty}
P\left[\frac{\gamma_{1}\gamma_{2}}{\gamma_{2}+C}
<\gamma \Big|\gamma_{1}\right]f_{\gamma_{1}}(\gamma_{1})d\gamma_{1}\nonumber\\
&=F_{\gamma_{1}}(\gamma)+\underbrace{\int_{\gamma}^{\infty}F_{\gamma_{2}}
\left(\frac{C\gamma}{\gamma_{1}-\gamma}\right)
f_{\gamma_{1}}(\gamma_{1})d\gamma_{1}}_{I_{1}},\tag{12}\label{12}
\end{align}
where $F_{\gamma_{1}}(\gamma)$ is the CDF of $\gamma_{1}$.
By substituting (\ref{6}), (\ref{7}) and (\ref{10}) into (\ref{12}), and after some algebraic manipulations,
we obtain
\begin{align}
&F_{\gamma^{AF}}(\gamma )=\frac{1}{\Gamma(m_w)\Gamma(k_w)}
\, {\rm {G}}_{1,3}^{2,1}\left [{\frac{\widetilde{\Xi}\gamma}{\overline\gamma_{1}}
\left |{ \begin{matrix} {1}\\ {k_w,m_w,0} \\ \end{matrix} }\right . }\right ]\!\nonumber\\
&+\frac{\omega r\widetilde{\Xi}\gamma}{\overline\gamma_{1}\Gamma(m_w)\Gamma(k_w)}\nonumber\\
& \times {\rm {H}_{1,0:3,1:2,1}^{0,1:1,2:0,2}} \left [{\!\!\left .{ \begin{matrix}
\left({2;-1,1}\right)\\ -\\ (0,r)(0,1)(1,r) \\ (0,r)\\ (2{-}k_w,1)(2{-}m_w,1)
\\ (1,1) \end{matrix} }\right |\!
\frac{\lambda^r \mu_{r}}{C}, \!\frac{\overline\gamma_{1}}{\widetilde{\Xi}\gamma} \!\!}\right ]\nonumber\\
&+\frac{(1{-}\omega) r\widetilde{\Xi}\gamma}{\overline\gamma_{1}\Gamma(m_w)\Gamma(k_w)\Gamma(a)c}\nonumber\\
& \times {\rm {H}_{1,0:3,1:2,1}^{0,1:1,2:0,2}} \left [{\!\!\left .{ \begin{matrix}
\left({2;-1,1}\right)\\ -\\ (1{-}a,\frac{r}{c})(0,1)(1,\frac{r}{c}) \\ (0,\frac{r}{c})
\\ (2{-}k_w,1)(2{-}m_w,1)\\ (1,1) \end{matrix} }\right |\!
\frac{b^r \mu_{r}}{C}, \!\frac{\overline\gamma_{1}}{\widetilde{\Xi}\gamma} \!\!}\right ],\tag{13}\label{13}
\end{align}
where $\rm {H}_{\cdot,\cdot:\cdot,\cdot:\cdot,\cdot}^{\cdot,\cdot:\cdot,\cdot:\cdot,\cdot}[\cdot,\cdot]$
is the Extended Generalized Bivariate Fox's H-Function (EGBFHF), which is also known as the Fox's H-function of two variables [37, Eq. (2.56)].

\textit{Proof:} See Appendix A.

A peculiar characteristic of this transcendental function is that many special functions can be formulated in terms of a Fox's H-function,
including the Extended Generalized Fivariate Meijer's G-Function (EGBMGF)
$\rm {G}_{\cdot,\cdot:\cdot,\cdot:\cdot,\cdot}^{\cdot,\cdot:\cdot,\cdot:\cdot,\cdot}[\cdot,\cdot]$,
defined in [36, Eq. (07.34.21.0081.01)]. It is known that the solution of many practical problems cannot be formulated in terms of traditional integral transforms, such as the Fourier, Laplace, and Mellin transforms. However, such problems can be efficiently characterized by integral transforms with specified functions as kernels \cite{39,40,41}. The EGBFHF, in particular, can be computed efficiently by, e.g., using the MATLAB implementation available in \cite{42}.

\begin{figure*}[b]
\hrulefill
\begin{align*}
&P_{out}^{AF,\infty} \approx\frac{\Gamma(|m_{w}-k_{w}|)}
{\Gamma(m_{w})\Gamma(k_{w})t}\left(\frac{\widetilde{\Xi}\gamma_{th}}{\overline{\gamma}_{1}}\right)^{t}
+\frac{w\Gamma\left(k_w-\frac{1}{r}\right)\Gamma\left(m_w-\frac{1}{r}\right)}
{\Gamma(m_{w})\Gamma(k_{w})}\left(\frac{C\widetilde{\Xi} \gamma_{th}}{\lambda^r\mu_{r}
\overline{\gamma}_{1}}\right)^{\frac{1}{r}}\nonumber\\
&+\frac{wr\Gamma\left(1-k_{w}r\right)\Gamma\left(m_w-k_w\right)\Gamma(k_{w}r)}
{\Gamma(m_{w})\Gamma(k_{w})\Gamma(1+k_{w}r)}\left(\frac{C\widetilde{\Xi}\gamma_{th}}{\lambda^r\mu_{r}
\overline{\gamma}_{1}}\right)^{k_w}+\frac{wr\Gamma\left(1-m_{w}r\right)
\Gamma\left(k_w-m_w\right)\Gamma(m_{w}r)}{\Gamma(m_{w})\Gamma(k_{w})\Gamma(1+m_{w}r)}
\left(\frac{C\widetilde{\Xi}\gamma_{th}}{\lambda^r\mu_{r}\overline{\gamma}_{1}}\right)^{m_w}\nonumber\\
&+\frac{(1-w)\Gamma\left(k_w-\frac{ac}{r}\right)\Gamma\left(m_w-\frac{ac}{r}\right)}{\Gamma(m_{w})\Gamma(k_{w})\Gamma(1+a)}
\left(\frac{C\widetilde{\Xi}\gamma_{th}}{b^r\mu_{r}\overline{\gamma}_{1}}\right)^{\frac{ac}{r}}
+\frac{(1-w)\Gamma\left(a-\frac{k_{w}r}{c}\right)\Gamma\left(m_{w}-k_{w}\right)
\Gamma\left(\frac{k_{w}r}{c}\right)r}{\Gamma(m_{w})\Gamma(k_{w})\Gamma(a)\Gamma\left(1+\frac{k_{w}r}{c}\right)c}
\left(\frac{C\widetilde{\Xi}\gamma_{th}}{b^r\mu_{r}\overline{\gamma}_{1}}\right)^{k_w}\nonumber\\
&+\frac{(1-w)\Gamma\left(a-\frac{m_{w}r}{c}\right)\Gamma\left(k_{w}-m_{w}\right)
\Gamma\left(\frac{m_{w}r}{c}\right)r}{\Gamma(m_{w})\Gamma(k_{w})\Gamma(a)\Gamma\left(1+\frac{m_{w}r}{c}\right)c}
\left(\frac{C\widetilde{\Xi}\gamma_{th}}{b^r\mu_{r}\overline{\gamma}_{1}}\right)^{m_w}.
\tag{22}\label{22}
\end{align*}
\end{figure*}

\subsubsection{Probability Density Function}
Taking the derivative of (\ref{12}), the PDF of the e2e SNR can be formulated as
\begin{align}
f_{\gamma^{AF}}(\gamma)=\frac{d}{d\gamma}\int_{0}^{\infty}
P\left[\frac{\gamma_{1}\gamma_{2}}{\gamma_{2}+C}
<\gamma \Big|\gamma_{1}\right]f_{\gamma_{1}}(\gamma_{1})d\gamma_{1}.\tag{14}\label{14}
\end{align}
Using the method in \cite{43}, and after some algebraic manipulations, the final expression of the PDF is
\begin{align}
f_{\gamma^{AF}}&(\gamma)=\frac{\omega \widetilde{\Xi}}
{\overline\gamma_{1}\Gamma(m_w)\Gamma(k_w)}\nonumber\\
&\times {\rm {H}_{1,0:2,0:2,1}^{0,1:0,2:0,2}} \left [{\!\!\left .{ \begin{matrix}
\left({2;-1,1}\right)\\ -\\ (0,r)(1,1) \\ -\\ (2{-}k_w,1)(2{-}m_w,1)\\ (2,1)
\end{matrix} }\right |\!\frac{\lambda^r \mu_{r}}{C}, \!
\frac{\overline\gamma_{1}}{\widetilde{\Xi}\gamma} \!\!}\right ]\nonumber\\
&+\frac{(1{-}\omega) \widetilde{\Xi}}{\overline\gamma_{1}\Gamma(m_w)\Gamma(k_w)\Gamma(a)}\nonumber\\
&\times {\rm {H}_{1,0:2,0:2,1}^{0,1:0,2:0,2}}
\left [{\!\!\left .{ \begin{matrix}\left({2;-1,1}\right)\\ -
\\ (1{-}a,\frac{r}{c})(1,1) \\ -\\ (2{-}k_w,1)(2{-}m_w,1)\\ (2,1) \end{matrix} }\right |\!
\frac{b^r \mu_{r}}{C}, \!\frac{\overline\gamma_{1}}{\widetilde{\Xi}\gamma} \!\!}\right ].\tag{15}\label{15}
\end{align}

\textit{Proof:} See Appendix B.

\begin{table}[t]
\centering
\caption{Parameters of the EGG Distribution for Different Bubble Levels BL (L/min)
for Fresh Water and Salty Water \cite{5}}
{\begin{tabular}{p{1.4cm}p{0.8cm}p{0.6cm}p{0.6cm}p{0.6cm}p{0.6cm}p{0.9cm}}
 \hline
 \hline
Salinity & BL (L/min) & $\omega$ &$\lambda$ & a & b & c\\
 \hline
Salty Water & 4.7 & 0.2064 & 0.3953 & 0.5307 & 1.2154 & 35.7368 \\
Salty Water & 7.1 & 0.4344 & 0.4747 & 0.3935 & 1.4506 & 77.0245 \\
Salty Water & 16.5 & 0.4951 & 0.1368 & 0.0161 & 3.2033 & 82.1030 \\
Fresh Water & 4.7 & 0.2190 & 0.4603 & 1.2526 & 1.1501 & 41.3258 \\
Fresh Water & 7.1 & 0.3489 & 0.4771 & 0.4319 & 1.4531 & 74.3650 \\
Fresh Water & 16.5 & 0.5117 & 0.1602 & 0.0075 & 2.9963 & 216.8356 \\
 \hline
 \hline
\end{tabular}}
\end{table}

\subsection{DF Relaying Protocol}
As for the DF relaying protocol, the e2e instantaneous SNR at D is defined as
\begin{align}
\gamma^{DF}=\mathrm{min}\{\gamma_{1},\gamma_{2}\}.\tag{16}\label{16}
\end{align}
\subsubsection{Cumulative Distribution Function}
The CDF of $\gamma^{DF}$ can be written as \cite{44}
\begin{align}
F_{\gamma^{DF}}(\gamma)&=\mathrm{Pr}(\mathrm{min}
\{\gamma_{1},\gamma_{2}\}<\gamma)\nonumber\\
&=F_{\gamma_{1}}(\gamma)+F_{\gamma_{2}}(\gamma)
-F_{\gamma_{1}}(\gamma)F_{\gamma_{2}}(\gamma),\tag{17}\label{17}
\end{align}
where $F_{\gamma_{1}}(\gamma)$ and $F_{\gamma_{2}}(\gamma)$ denote
the CDFs of $\gamma_{1}$ and $\gamma_{2}$, respectively. By substituting (\ref{7}) and (\ref{10}) in (\ref{17}), the CDF can be formulated in closed-form.

\subsubsection{Probability Density Function}
By taking the derivative of (\ref{17}) with respect to $\gamma$,
the following PDF is obtained
\begin{align}
f_{\gamma^{DF}}(\gamma)=&f_{\gamma_{1}}(\gamma)+f_{\gamma_{2}}(\gamma)
-f_{\gamma_{1}}(\gamma)F_{\gamma_{2}}(\gamma)\nonumber\\
&-f_{\gamma_{2}}(\gamma)F_{\gamma_{1}}(\gamma),\tag{18}\label{18}
\end{align}
which can be obtained from the PDFs and CDFs of $\gamma_1$ and $\gamma_2$.

\section{Performance Analysis}
In this section, exact closed-form expressions for the OP, ABER, and ACC are derived assuming fixed-gain AF and DF relays. Moreover, in order to have a better understanding of the impact of different system and fading parameters on the e2e performance, tight asymptotic expressions in the high SNR regime are derived, based on which the diversity order is computed.

\begin{figure*}[b]
\hrulefill
\begin{align*}
\overline{P}_{out}^{AF,\infty} \approx &\frac{\Gamma(|m_{w}-k_{w}|)\Gamma(p+t)}{2\Gamma(p)\Gamma(m_{w})\Gamma(k_{w})t}
\left(\frac{\widetilde{\Xi}}{\overline{\gamma}_{1}q}\right)^{t}+\frac{w\Gamma\left(k_w-\frac{1}{r}\right)
\Gamma\left(m_w-\frac{1}{r}\right)\Gamma\left(p+\frac{1}{r}\right)}{2\Gamma(p)\Gamma(m_{w})\Gamma(k_{w})}
\left(\frac{C\widetilde{\Xi}}{\lambda^rq\mu_{r}\overline{\gamma}_{1}}\right)^{\frac{1}{r}}\nonumber\\
&+\frac{wr\Gamma\left(1-k_{w}r\right)\Gamma\left(m_w-k_w\right)\Gamma(k_{w}r)\Gamma(p+k_w)}
{2\Gamma(p)\Gamma(m_{w})\Gamma(k_{w})\Gamma(1+k_{w}r)}\left(\frac{C\widetilde{\Xi}}{\lambda^rq\mu_{r}
\overline{\gamma}_{1}}\right)^{k_w}\nonumber\\
&+\frac{wr\Gamma\left(1-m_{w}r\right)\Gamma\left(k_w-m_w\right)\Gamma(m_{w}r)\Gamma(p+m_w)}
{2\Gamma(p)\Gamma(m_{w})\Gamma(k_{w})\Gamma(1+m_{w}r)}\left(\frac{C\widetilde{\Xi}}{\lambda^rq\mu_{r}
\overline{\gamma}_{1}}\right)^{m_w}\nonumber\\
&+\frac{(1-w)\Gamma\left(k_w-\frac{ac}{r}\right)\Gamma\left(m_w-\frac{ac}{r}\right)
\Gamma\left(p+\frac{ac}{r}\right)}{2\Gamma(p)\Gamma(m_{w})\Gamma(k_{w})\Gamma(1+a)}
\left(\frac{C\widetilde{\Xi}}{b^rq\mu_{r}\overline{\gamma}_{1}}\right)^{\frac{ac}{r}}\nonumber\\
&+\frac{(1-w)\Gamma\left(a-\frac{k_{w}r}{c}\right)\Gamma\left(m_{w}-k_{w}\right)\Gamma\left(\frac{k_{w}r}{c}\right)
\Gamma(p+k_w)r}{2\Gamma(p)\Gamma(m_{w})\Gamma(k_{w})\Gamma(a)\Gamma\left(1+\frac{k_{w}r}{c}\right)c}
\left(\frac{C\widetilde{\Xi}}{b^rq\mu_{r}\overline{\gamma}_{1}}\right)^{k_w}\nonumber\\
&+\frac{(1-w)\Gamma\left(a-\frac{m_{w}r}{c}\right)\Gamma\left(k_{w}-m_{w}\right)
\Gamma\left(\frac{m_{w}r}{c}\right)\Gamma(p+m_w)r}{2\Gamma(p)\Gamma(m_{w})\Gamma(k_{w})\Gamma(a)\Gamma
\left(1+\frac{m_{w}r}{c}\right)c}\left(\frac{C\widetilde{\Xi}}{b^rq\mu_{r}\overline{\gamma}_{1}}\right)^{m_w}.
\tag{28}\label{28}
\end{align*}
\end{figure*}

\subsection{Fixed-Gain AF Relaying}
\subsubsection{Outage Probability}
The OP, $P_{out}$, is defined as the probability that the instantaneous SNR $\gamma$ falls below a
given threshold $\gamma_{th}$. Therefore, the exact OP expression of the considered system with a
fixed-gain AF relay can be easily obtained from (\ref{13}), i.e.,
\begin{align}
P_{out}^{AF}=\Pr[\gamma<\gamma_{th}]=F_{\gamma^{AF}}(\gamma_{th}).\tag{19}\label{19}
\end{align}

Since (\ref{19}) is written in terms of EGBFHF, it is hard to get engineering insights
from it. Therefore, to obtain more useful insights, an asymptotic analysis is carried out by setting $\overline{\gamma}_1=\overline{\gamma}_2=\overline{\gamma}\rightarrow \infty$, which yields
\begin{align}
P_{out}^{AF,\infty}\approx F_{\gamma_{1}}^{\infty}+I_{1}^{\infty},\tag{20}\label{20}
\end{align}
where $F_{\gamma_{1}}^{\infty}$ is the asymptotic OP of $\gamma_{1}$ and $I_{1}^{\infty}$ is the asymptotic
result of $I_{1}$ given in (\ref{12}). By using [36, Eq. (07.34.06.0040.01)], $F_{\gamma_{1}}^{\infty}$
can be formulated as
\begin{align}
F_{\gamma_{1}}^{\infty}(\gamma_{th})=\frac{\Gamma(|m_w-k_w|)
}{\Gamma(k_w)\Gamma(m_w)t}\left(\frac{E_w\gamma_{th}}{\overline\gamma_{1}}\right)^{t},\tag{21}\label{21}
\end{align}
where $t=\min(m_w,k_w)$. From [45, Eq. (1.1)], [46, Th. 1.11] and [46, Eq. (1.8.4)], the
OP of fixed-gain AF relaying, $P_{out}^{AF,\infty}$, in the high SNR regime is given in (\ref{22}), shown at the bottom of this page. Note that the asymptotic expression in (\ref{22}) is written in terms of elementary functions that can be computed by using any computer software.

\textit{Proof:} See Appendix C.

From (\ref{22}), in addition, we evince that, the diversity order of the considered system setup is
\begin{align}
G_{d}^{AF} = \min\left(m_w,k_w,\frac{2}{r},\frac{2ac}{r}\right).\tag{23}\label{23}
\end{align}
From (\ref{23}), we observe that the diversity order is a function of $m_{w}$, $k_w$, $a$, $c$, and $r$.
Based on Section II, in addition, $m_{w}$ and $k_w$ depend on $N$ and $m$.
Therefore, the diversity order is determined by $m$, $N$, and by the detection technique (i.e., $r$) employed.

\subsubsection{Average Bit Error Rate}
The ABER of several binary modulation schemes can be expressed as [44, Eq. (25)]
\begin{align}
\overline{P}_{e}=\frac{q^p}{2\Gamma(p)}\int_{0}^{\infty}\exp(-q\gamma)
\gamma^{p-1}F_{\gamma}(\gamma)d\gamma,\tag{24}\label{24}
\end{align}
where $p$ and $q$ depend on the modulation scheme. For example, binary phase shift keying (BPSK) is obtained by setting $p=0.5$ and $q=1$.
Then, by substituting (\ref{12}) into (\ref{24}), the ABER of fixed-gain AF relaying can be formulated as
\begin{align}
&\overline{P}_{e}^{AF}{=}\frac{q^p}{2\Gamma(p)}\int_{0}^{\infty}\gamma^{p{-}1}
\exp({-}q\gamma)F_{\gamma_{1}}(\gamma)d\gamma{+}\frac{q^p}{2\Gamma(p)}\nonumber\\
&{\times}\int_{0}^{\infty}\gamma^{p{-}1}\exp(-q\gamma)\int_{\gamma}^{\infty}F_{\gamma_{2}}
\left(\frac{C\gamma}{\overline{\gamma}_1{-}\gamma}\right)f_{\gamma_{1}}(\gamma_{1})d\gamma_{1}d\gamma\nonumber\\
&=\overline{P}_{e,1}+I_{2},\tag{25}\label{25}
\end{align}
where $\overline{P}_{e,1}$ is the ABER of the RF link. By replacing (\ref{7}) into (\ref{24}), expressing $\exp(-q\gamma)$ in terms of Meijer's G-function [35, Eq. (8.4.3.1)], and utilizing [36, Eq. (07.34.21.0011.01)], we obtain
\begin{align}
\overline{P}_{e,1}=\frac{1}{2\Gamma(p)\Gamma(m_w)\Gamma(k_w)}\, {\mathrm{G}}_{2,3}^{2,2}
\left [{{\frac{\widetilde{\Xi}}{q\overline\gamma_{1}} }\left |{ \begin{matrix} {1,1{-}p}
\\ {k_w,m_w,0} \\ \end{matrix} }\right . }\right ]\!.\tag{26}\label{26}
\end{align}
Similar to (\ref{13}), $I_{2}$ can be formulated as
\begin{align}
&I_{2}=\frac{\omega r\widetilde{\Xi}}{2\overline\gamma_{1}\Gamma(p)\Gamma(m_w)\Gamma(k_w)q}\nonumber\\
&\times {\rm {H}_{1,0:3,1:2,2}^{0,1:1,2:1,2}} \left [{\!\!\left .{ \begin{matrix}
\left({2;-1,1}\right)\\ -\\ (0,r)(0,1)(1,r) \\ (0,r)\\ (2{-}k_w,1)(2{-}m_w,1)\\ (1{+}p,1)(1,1) \end{matrix} }\right |\!
\frac{\lambda^r \mu_{r}}{C}, \!\frac{q\overline\gamma_{1}}{\widetilde{\Xi}} \!\!}\right ]\nonumber\\
&+\frac{(1{-}\omega) r\widetilde{\Xi}}{2\overline\gamma_{1}\Gamma(p)\Gamma(m_w)\Gamma(k_w)\Gamma(a)cq}\nonumber\\
&\times {\rm {H}_{1,0:3,1:2,2}^{0,1:1,2:1,2}} \left [{\!\!\left .{ \begin{matrix}
\left({2;-1,1}\right)\\ -\\ (1{-}a,\frac{r}{c})(0,1)(1,\frac{r}{c}) \\ (0,\frac{r}{c})\\ (2{-}k_w,1)(2{-}m_w,1)\\ (1{+}p,1)(1,1) \end{matrix} }\right |\!
\frac{b^r \mu_{r}}{C}, \!\frac{q\overline\gamma_{1}}{\widetilde{\Xi}} \!\!}\right ].
\tag{27}\label{27}
\end{align}

\textit{Proof:} Please, see Appendix D.

By letting $\overline{\gamma}_1=\overline{\gamma}_2=\overline{\gamma}\rightarrow \infty$ and employing
the same method presented in Appendix C, the asymptotic ABER can be obtained as formulated in
(\ref{28}), shown at the bottom of this page. Similar to the asymptotic outage analysis, the diversity order calculated through the
asymptotic ABER coincide, as expected, with (23).

\subsubsection{Average Channel Capacity}
The ACC of the considered system setup operating under both HD and IM/DD can be formulated as [38, Eq. (25)] [47, Eq. (15)]
\begin{align}
\overline{C}=\frac{1}{2\ln(2)}\int_{0}^{\infty}
\ln(1+\tau\gamma)f_{\gamma^{AF}}(\gamma)d\gamma,\tag{29}\label{29}
\end{align}
where $\tau$ is a constant such that $\tau=e/(2\pi)$ holds for the IM/DD technique (i.e., when $r=2$) and
$\tau=1$ holds for the HD technique (i.e., when $r=1$) \cite{38,44}. It is worth noting that the expression in
(\ref{29}) is exact if $r=1$ while it is a lower-bound if $r=2$ \cite{47}.
Using [35, Eq. (8.4.6.5)] and [36, Eq. (07.34.26.0008.01)], $\ln(1+\tau\gamma)$ can be represented as
\begin{align}
\ln(1+\tau\gamma)=\, {\rm {H}}_{2,2}^{1,2}\left [{{\tau\gamma}\left |
{ \begin{matrix} {(1,1)(1,1)}\\{(1,1)(0,1)} \\ \end{matrix} }\right . }\right ]\!.\tag{30}\label{30}
\end{align}
By inserting (\ref{15}) and (\ref{30}) into (\ref{29}), and then applying [37, Eqs. (2.8) and (2.57)],
the ACC of fixed-gain AF relaying can be derived in closed-form as
\begin{align}
&\overline{C}^{AF}=\frac{\omega \widetilde{\Xi}}{2\ln(2)\tau
\overline\gamma_{1}\Gamma(m_w)\Gamma(k_w)}\nonumber\\
&\times {\rm {H}_{1,0:2,0:3,2}^{0,1:0,2:1,3}}
\left [{\!\!\left .{ \begin{matrix}
\left({2;-1,1}\right)\\ -\\ (0,r)(1,1) \\ -
\\ (2{-}k_w,1)(2{-}m_w,1)(2,1)\\ (2,1)(1,1)
\end{matrix} }\right |\! \frac{\lambda^r \mu_{r}}{C},
\!\frac{\overline\gamma_{1}\tau}{\widetilde{\Xi}} \!\!}\right ]\nonumber\\
&+\frac{(1-\omega) \widetilde{\Xi}}{2\ln(2)\tau\overline\gamma_{1}
\Gamma(m_w)\Gamma(k_w)\Gamma(a)}\nonumber\\
&\times {\rm {H}_{1,0:2,0:3,2}^{0,1:0,2:1,3}}
\left [{\!\!\left .{ \begin{matrix}
\left({2;-1,1}\right)\\ -\\ (1{-}a,\frac{r}{c})(1,1)
\\ -\\ (2{-}k_w,1)(2{-}m_w,1)(2,1)\\ (2,1)(1,1)
\end{matrix} }\right |\! \frac{b^r \mu_{r}}{C},
\!\frac{\overline\gamma_{1}\tau}{\widetilde{\Xi}} \!\!}\right ].\tag{31}\label{31}
\end{align}
When $m=1$, $r=1$, and $c=1$, (\ref{31}) reduces to a special case where both the S-RIS and RIS-R links
undergo Rayleigh fading, while the UWOC link undergoes EG fading and employs the HD technique.


\subsection{DF Relaying}
\subsubsection{Outage Probability}
As for the DF relaying, the OP is obtained from (17) as
\begin{align}
P_{out}^{DF}=F_{\gamma^{DF}}(\gamma_{th}).\tag{32}\label{32}
\end{align}

Then, the asymptotic CDF of $\gamma^{DF}$ can be expressed as
\begin{align}
P_{out}^{DF,\infty}&=F_{\gamma_{1}}^{\infty}(\gamma_{th})
+F_{\gamma_{2}}^{\infty}(\gamma_{th})
-F_{\gamma_{1}}^{\infty}(\gamma_{th})
F_{\gamma_{2}}^{\infty}(\gamma_{th})\nonumber\\
&\approx F_{\gamma_{1}}^{\infty}(\gamma_{th})
+F_{\gamma_{2}}^{\infty}(\gamma_{th}),\tag{33}\label{33}
\end{align}
where $F_{\gamma_{1}}^{\infty}(\cdot)$ is given in (\ref{21}). When $\mu_{r}\rightarrow \infty$,
$F_{\gamma_{2}}^{\infty}(\gamma_{th})$ can be simplified by using [46, Eq. (1.8.4)], which yields
\begin{align}
F_{\gamma_{2}}^{\infty}(\gamma_{th})=\omega
\left(\frac{\gamma_{th}}{\lambda^r \mu_{r}}\right)^{\frac{1}{r}}
+\frac{(1-\omega)}{\Gamma(a+1)}
\left(\frac{\gamma_{th}}{b^r \mu_{r}}\right)^{\frac{ac}{r}},\tag{34}\label{34}
\end{align}
Therefore, by using (\ref{21}) and (\ref{34}), the diversity order is
\begin{align}
G_{d}^{DF} = \min\left(m_w,k_w,\frac{1}{r},\frac{ac}{r}\right).\tag{35}\label{35}
\end{align}
Similar to the AF relaying, the diversity order of the DF
relaying is determined by $m$, $N$, and by the detection scheme.

\subsubsection{Average Bit Error Rate}
As for the DF relaying, the ABER can be expressed as
\begin{align}
\overline{P}_{e}^{DF}&=\overline{P}_{e,1}+\overline{P}_{e,2}-2\overline{P}_{e,1}\overline{P}_{e,2},\tag{36}\label{36}
\end{align}
where $\overline{P}_{e,1}$ and $\overline{P}_{e,2}$ are the ABER of the RF and UWOC links, respectively. In particular, $\overline{P}_{e,1}$ is
given in (\ref{26}). By substituting (\ref{10}) into (\ref{24}) and applying the identity $\exp(-q\gamma){=}
\, {\rm {H}}_{0,1}^{1,0}\left [{q\gamma \left |{ \begin{matrix} {-}
\\ {(0,1)} \\ \end{matrix} }\right . }\right ]\!$ [35, Eq. (8.4.3.1)], [36, Eq. (07.34.26.0008.01)],
and then utilizing [35, Eq. (2.25.1.1)], we obtain
\begin{align}
\overline{P}_{e,2}&=\frac{\omega r}{2\Gamma(p)}\, {\rm {H}}_{2,2}^{1,2}
\left [{{\frac{1}{\lambda^r\mu_{r}q}}\left |{ \begin{matrix} {(1,r)(1{-}p,1)}
\\ {(1,r)(0,r)} \\ \end{matrix} }\right . }\right ]\! \nonumber\\
&+\frac{(1-\omega)r}{2\Gamma(p)\Gamma(a)c}\, {\rm {H}}_{2,2}^{1,2}
\left [{{\frac{1}{b^r\mu_{r}q}}\left |{ \begin{matrix} {(1,\frac{r}{c})(1{-}p,1)}
\\ {(a,\frac{r}{c})(0,\frac{r}{c})} \\ \end{matrix} }\right . }\right ]\!.\tag{37}\label{37}
\end{align}
Finally, from (\ref{26}), (\ref{36}) and (\ref{37}), a closed-form expression for the
ABER can be derived. It is noteworthy that an efficient MATHEMATICA implementation for evaluating the Fox's H-function can be found in \cite{48}.

By using [36, Eq. (07.34.06.0040.01)] and [46, Eq. (1.8.4)], an asymptotic expression for the ABER is given as
\begin{align}
\overline{P}_{e}^{DF,\infty}\approx \overline{P}_{e,1}^{\infty}+\overline{P}_{e,1}^{\infty},\tag{38}\label{38}
\end{align}
where
\begin{align}
\overline{P}_{e,1}^{\infty}=\frac{\Gamma(|m_w-k_w|)\Gamma(p+t)}
{2\Gamma(p)\Gamma(k_w)\Gamma(m_w)t}\left(\frac{E_w}{\overline\gamma_{1}q}\right)^{t},\tag{39}\label{39}
\end{align}
and
\begin{align}
\overline{P}_{e,2}^{\infty}=\frac{\omega}{2\Gamma(p)}
&\Gamma\left(p+\frac{1}{r}\right)\left(\frac{1}{\lambda^r \mu_{r}q}\right)^{\frac{1}{r}}\nonumber\\
&+\frac{(1-\omega)\Gamma\left(p+\frac{ac}{r}\right)}{2\Gamma(p)\Gamma(a+1)}
\left(\frac{1}{b^r \mu_{r}q}\right)^{\frac{ac}{r}}.\tag{40}\label{40}
\end{align}

\subsubsection{Average Channel Capacity}
From (\ref{29}) and using (\ref{18}) and (\ref{30}), the ACC can be formulated as
\begin{align}
\overline{C}^{DF}=&\frac{1}{2\ln(2)}\int_{0}^{\infty}\, {\rm {H}}_{2,2}^{1,2}
\left [{{\tau\gamma}\left |{ \begin{matrix} {(1,1)(1,1)}\\{(1,1)(0,1)} \\
\end{matrix} }\right . }\right ]\!f_{\gamma}^{DF}(\gamma)d\gamma\nonumber\\
=&I_{C1}+I_{C2}-I_{C3}-I_{C4}.\tag{41}\label{41}
\end{align}
Then, by applying [35, Eq. (2.25.1.1)] and after some algebraic operations, we have
\begin{align}
I_{C1}&=\frac{\widetilde{\Xi}}{2\ln(2)\overline\gamma_{1}\Gamma(m_w)\Gamma(k_w)\tau}\nonumber\\
\times & \, {\rm {H}}_{2,4}^{4,1}\left [{{\frac{\widetilde{\Xi}}{\overline\gamma_{1}\tau}}
\left |{ \begin{matrix} {(-1,1)(0,1)}\\{(k_w{-}1,1)(m_w{-}1,1)(-1,1)(-1,1)}
\\ \end{matrix} }\right . }\right ]\!.\tag{42}\label{42}
\end{align}
Similarly, $I_{C2}$ is derived as
\begin{align}
&I_{C2}=\frac{\omega r}{2\ln(2)}\, {\rm {H}}_{1,2}^{2,1}
\left [{{\frac{1}{\lambda^r\mu_{r}\tau}}\left |{ \begin{matrix} {(0,1)}
\\ {(0,r)(0,1)} \\ \end{matrix} }\right . }\right ]\!\nonumber\\
&+\frac{(1-\omega) r}{2\ln(2)\Gamma(a)c}\, {\rm {H}}_{2,3}^{3,1}
\left [{{\frac{1}{b^r\mu_{r}\tau}}\left |{ \begin{matrix} {(0,1)(1,\frac{r}{c})}
\\ {(a,\frac{r}{c})(0,\frac{r}{c})(0,1)}
\\ \end{matrix} }\right . }\right ]\!.\tag{43}\label{43}
\end{align}
In addition, from (\ref{6}) and (\ref{10}), and utilizing [36, Eq. (07.34.26.0008.01)] and
[45, Eqs. (1.1) and (2.3)], $I_{C3}$ can be formulated as
\begin{align}
&I_{C3}{=}\frac{wr}{2\ln(2)\Gamma(m_w)\Gamma(k_w)}\nonumber\\
&\times {\rm {H}_{2,0:2,2:1,2}^{0,2:1,2:1,1}}
\left [{\!\!\left .{ \begin{matrix} (1{-}k_w;1,1)(1{-}m_w;1,1)\\ -\\ (1,1)(1,1)
\\ (1,1)(0,1)\\ (1,r)\\ (1,r)(0,r) \end{matrix} }\right |\!
\frac{\overline\gamma_{1}\tau}{\widetilde{\Xi}},
\!\frac{\overline\gamma_{1}}{\lambda^r\mu_{r}\widetilde{\Xi}} \!\!}\right ]\nonumber\\
&{+}\frac{(1-w)r}{2\ln(2)\Gamma(m_w)\Gamma(k_w)\Gamma(a)}\nonumber\\
& \times {\rm {H}_{2,0:2,2:1,2}^{0,2:1,2:1,1}} \left [{\!\!
\left .{ \begin{matrix} (1{-}k_w;1,1)(1{-}m_w;1,1)\\ -\\ (1,1)(1,1) \\ (1,1)(0,1)\\ (1,\frac{r}{c})\\
(a,\frac{r}{c})(0,\frac{r}{c}) \end{matrix} }\right |\! \frac{\overline\gamma_{1}\tau}{\widetilde{\Xi}}, \!
\frac{\overline\gamma_{1}}{b^r\mu_{r}\widetilde{\Xi}} \!\!}\right ],\tag{44}\label{44}
\end{align}
while with the help of (\ref{7}), (\ref{9}) and (\ref{41}), $I_{C4}$ is given as follows
\begin{align}
&I_{C4}=\frac{w}{2\ln(2)\Gamma(m_w)\Gamma(k_w)}\nonumber\\
& \times {\rm {H}_{1,0:2,2:1,3}^{0,1:1,2:2,1}}
\left [{\!\!\left .{ \begin{matrix} (0;r,r)\\ -\\ (1,1)(1,1)
\\ (1,1)(0,1)\\ (1,1)\\ (k_w,1)(m_w,1)(0,1)
\end{matrix} }\right |\! \tau \lambda^r\mu_{r}, \!
\frac{\lambda^r\mu_{r}\widetilde{\Xi}} {\overline\gamma_{1}} \!\!}\right ]\nonumber\\
&{+}\frac{(1-w)}{2\ln(2)\Gamma(m_w)\Gamma(k_w)\Gamma(a)}\nonumber\\
& \times {\rm {H}_{1,0:2,2:1,3}^{0,1:1,2:2,1}}
\left [{\!\!\left .{ \begin{matrix} (1{-}a;\frac{r}{c},\frac{r}{c})
\\ -\\ (1,1)(1,1) \\ (1,1)(0,1)\\ (1,1)\\ (k_w,1)(m_w,1)(0,1)
\end{matrix} }\right |\! \tau b^r\mu_{r},
\!\frac{b^r\mu_{r}\widetilde{\Xi}}{\overline\gamma_{1}} \!\!}\right ].\tag{45}\label{45}
\end{align}

By setting $r=1$ and $c=1$, as a special case for the DF relaying, (\ref{45}) is simplified to the setup when
the UWOC link follows a EG distribution and employs the HD technology.

\section{Numerical Results and Discussions}
In this section, illustrative numerical examples are provided in order to investigate the impact of key system parameters on the e2e system performance. Without loss of generality, we set the threshold $\gamma_{th}=2$ dB, the fading parameter $m=2$, and the fixed relay
gain $C=1.5$. The parameters of the UWOC link are given in Table I and Table II. As for the system without an RIS,
we assume that S is equipped with one antenna and communicates directly with R over a Rayleigh fading channels.
As it will be observed from the figures, the simulation results match perfectly with the analytical results, thus confirming the
accuracy of our analytical derivation.

\begin{figure}[t]
    \centering
    \includegraphics[width=3.5in]{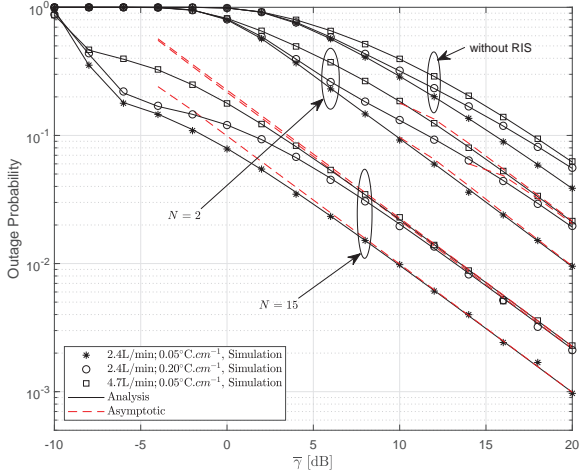}
    \caption{OP versus $\overline{\gamma}$ for fixed-gain AF relaying under the IM/DD technique and by varying $N$.}
\end{figure}

\begin{figure}[t]
    \centering
    \includegraphics[width=3.5in]{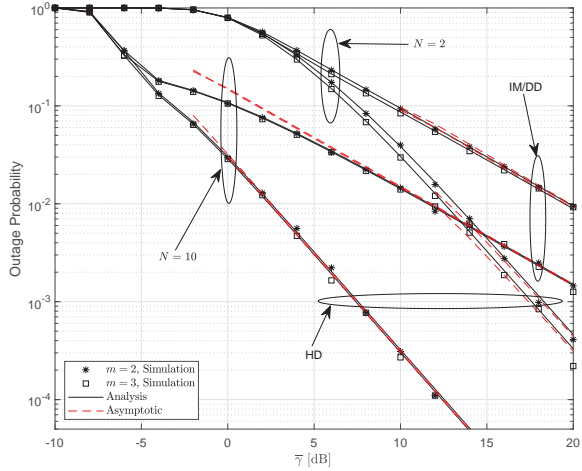}
    \caption{OP versus $\overline{\gamma}$ for fixed-gain AF relaying under both HD and IM/DD technique and by setting $m=2,3$.}
\end{figure}

Fig. 2 shows the OP versus $\overline{\gamma}$ of the considered fixed-gain AF relay-aided system with and without the RIS, where $\overline{\gamma}$ is the average SNRs of both links, i.e., $\overline{\gamma}_1=\overline{\gamma}_2=\overline{\gamma}$. Moreover, we assume that the system employs the IM/DD technique and operates under variable turbulence conditions. Compared to the scheme without the RIS, one can observe that the system outage performance in the presence of an RIS is improved. In particular, the performance the system outage performance gets better with $N$. In addition, in Table I, the higher the level of the air bubbles and/or the temperature gradient is, the higher the intensity of turbulence is. Therefore, it can be seen from Fig. 2 that the bubble level and the temperature gradient affect the performance of the system. As the level of the air bubbles and temperature gradients increase, the turbulence conditions are more severe, thereby resulting in performance deterioration. Furthermore, compared with the setup $\mathrm{BL}=2.4$ L/min, $\Delta \mathrm{T}=0.05 ^{\circ}\mathrm{C}.cm^{-1}$, the outage performance under the other two turbulent conditions shows that the effect of the level of air bubbles is more severe than the level of the temperature gradient. The asymptotic results derived in (\ref{22}) are in very tight agreement with the exact analysis in the high SNR regime.

\begin{figure}[t]
    \centering
    \includegraphics[width=3.5in]{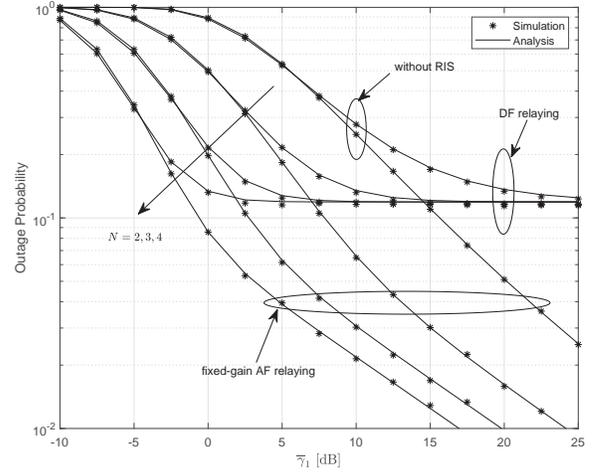}
    \caption{OP versus $\overline{\gamma}_{1}$ for both fixed-gain AF relaying and DF relaying under IM/DD technique.}
\end{figure}

\begin{figure}[t]
    \centering
    \includegraphics[width=3.5in]{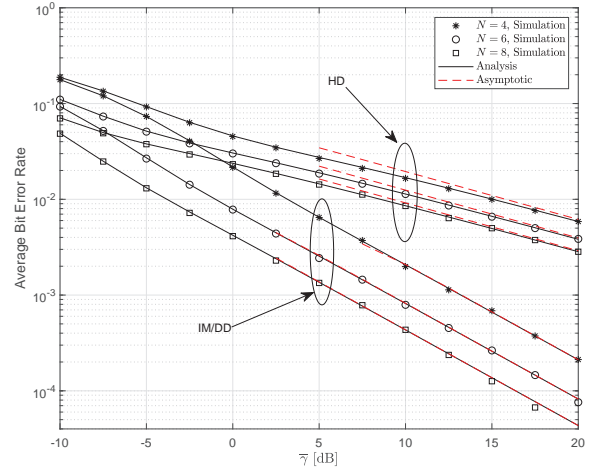}
    \caption{ABER versus $\overline{\gamma}$ for fixed-gain AF relaying under both HD and IM/DD techniques.}
\end{figure}

Fig. 3 depicts the system outage performance versus $\overline{\gamma}$ by considering both HD and IM/DD techniques and by setting $m=2,3$.
This figure demonstrates that the HD technique provides better outage performance than the IM/DD technique.
This improvement in performance is due to the fact that the HD technique can better overcome the
turbulence effect compared to the IM/DD technique. When $N$ is small, the system provides better outage
performance if $m=3$. This is because a large value of $m$ indicates weaker fading
on the RF link. When $N$ is large, the parameter $m$ has little effect on the system performance. This
is because with the help of an RIS, the increase of $N$ reduces the OP of the RF link. Therefore, at
high SNR, the OP of the overall system is mainly determined by the UWOC link. Furthermore, we observe that
the curves have the same slopes when the system utilizes the same detection technology.
Therefore, it can be observed that the diversity order depends on the detection technique
of the UWOC link, which is consistent with (\ref{23}).

In Fig. 4, we plot the OP versus $\overline{\gamma}_1$ for both fixed-gain AF and DF
relays and by considering the IM/DD technique. We assume $\overline{\gamma}_{2}=15$ dB. As expected, the OP of the DF relaying
is worse than the OP of fixed-gain AF relaying system. For large values of $ \overline{\gamma}_1$, in particular, the OP of DF relaying is mainly determined by the UWOC link, and, therefore, it is almost independent of the RF link. The
outage performance improves by using an RIS and the OP decreases as $\overline{\gamma}_{1}$ and $N$ increase.

\begin{figure}[t]
    \centering
    \includegraphics[width=3.5in]{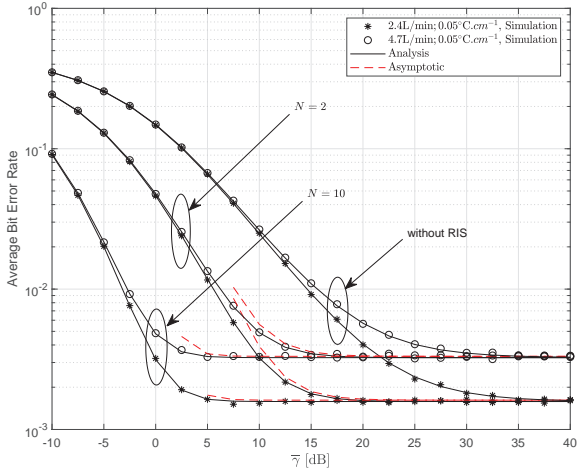}
    \caption{ABER versus $\overline{\gamma}$ for DF relaying under various levels of air bubbles and gradient temperatures.}
\end{figure}

\begin{figure}[t]
    \centering
    \includegraphics[width=3.5in]{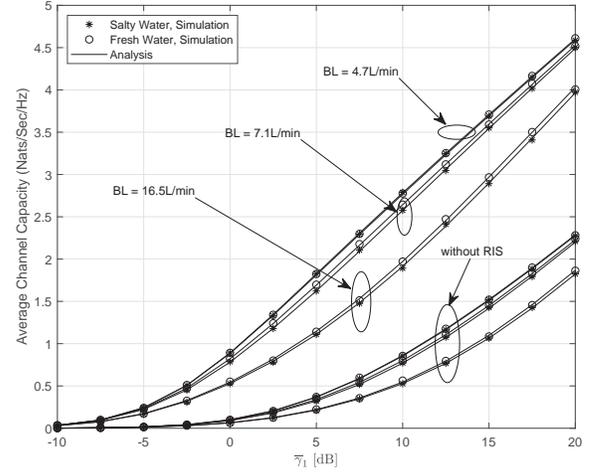}
    \caption{Average channel capacity versus $\overline{\gamma}_1$ for fixed-gain AF relaying and considering
    three different levels of air bubbles using both salty and fresh waters in the case of the IM/DD technique.}
\end{figure}

Fig. 5 depicts the ABER versus $\overline{\gamma}$ for fixed-gain AF relaying and for both HD and IM/DD detection
techniques. It is assumed that the turbulence condition corresponds to $\mathrm{BL}=2.4$ L/min,
$\Delta \mathrm{T}=0.05 ^{\circ}\mathrm{C}.cm^{-1}$. In this setup, the ABER performance
improves as $N$ increases. As expected, the HD technique provides better error performance than the IM/DD technique.
The asymptotic results of the ABER at high SNR are also shown. The asymptotic results are
in a perfect agreement with the analytical results in the high SNR regime. This justifies the accuracy and tightness of (\ref{28}).

In Fig. 6, the ABER versus $\overline{\gamma}_{1}$ is depicted for DF relaying and the HD technique by considering various levels of air bubbles and gradient temperatures. The error performance of the proposed system is better than the system without RISs, as expected. For systems with the same number of reflecting elements, the error performance is improved under weak turbulence conditions. In general, the larger the number of reflecting elements is the lower the ABER is.
In addition, under a given type of turbulence, for a fixed $\overline{\gamma}_{1}$, the ABER tends to be constant when $\overline{\gamma}_{2}=\infty$. This is because for a fixed $\overline{\gamma}_{1}$ and a large $\overline{\gamma}_{2}$, the overall system SNR $\gamma^{DF}$ is determined by $\gamma_{1}$ due to the fact that $\gamma^{DF}=\min\{\gamma_{1},\gamma_{2}\}$.
Furthermore, the asymptotic ABER in (\ref{38}) is in agreement with (\ref{36}) in the high SNR regime.

\begin{figure}[t]
    \centering
    \includegraphics[width=3.5in]{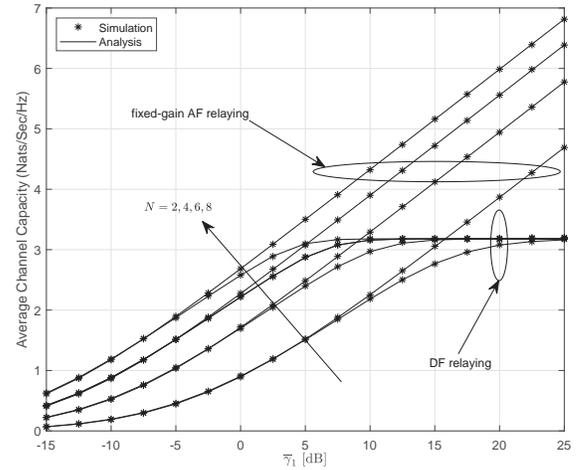}
    \caption{Average channel capacity versus $\overline{\gamma}_1$ for both fixed-gain AF and DF relaying and assuming
    HD technique.}
\end{figure}

Fig. 7 shows the ACC versus $\overline{\gamma}_1$ for fixed-gain AF relaying and assuming different levels of air bubbles, for both salty and fresh waters, and considering the IM/DD technique. We assume $N=5$. We observe that, for a given type of water, the ACC decreases as the turbulence conditions are more severe. For instance, when the levels of air bubbles increase from 4.7 L/min to 16.5 L/min, the ACC decrease to a greater extent. The effect of water salinity is analyzed in Fig. 7. We note that its impact is less significant than the impact of air bubbles. Under the assumption of weak turbulence intensity, e.g., BL$=4.7$ L/min, the impact of water salinity is almost negligible. In addition, we observe that the ACC of a system without RIS is always worse than that of a system with RIS.

Assuming the HD technique, Fig. 8 shows the ACC versus $\overline{\gamma}_1$ for both fixed-gain AF and DF relaying, and for different values of $N$. We assume $\overline{\gamma}_{2}=20$ dB. As for DF relaying, the ACC reaches an a asymptote for large values of $\overline{\gamma}_1$, which is similar as the observations made in Fig. 4. Clearly, this figure also demonstrates the ACC of the fixed-gain AF relaying is better than the ACC of DF relaying.

\section{Conclusion}
In this work, we investigated the performance of RIS-assisted dual-hop mixed RF-UWOC systems. Under different
detection techniques, exact closed-form expressions for the OP, ABER, and ACC were formulated in terms of Meijer's
G-function and Fox's H-function for both fixed-gain AF and DF relaying schemes. The impact of RISs on the system
performance was investigated and it was shown that an RIS can effectively improve the overall system performance.
Also, when $N$ or $m$ reaches a large value, the performance of the considered system setup mainly depends on the
UWOC link. Furthermore, we have derived tight asymptotic results for the obtained performance metrics, which are
written in terms of simple elementary functions. The system diversity order was determined and it is shown to be
determined by the RF link parameters (including $m$ and $N$), and by the detection technology of the UWOC link.
Furthermore, the impact of temperature gradients, air bubbles, fresh and salt water
under different turbulent conditions were analyzed. Our results demonstrated that the performance in the presence of lower bubbles level and temperature gradient is better than that with a higher bubbles level and temperature gradient. Moreover, it was shown that the water salinity has an effect on the performance of the mixed RF-UWOC system, but its effect is less pronounced than that of the air bubbles. If the bubble level is high, the performance in fresh water is better than in salt water. Compared to a DF relaying, an RIS-assisted dual-hop mixed RF-UWOC system with a fixed-gain AF relay attains better performance.

\appendices
\section{CDF of The End-to-End SNR for Fixed-Gain AF Relaying}
From (\ref{12}), performing the change of variable $x=\gamma_{1}-\gamma$,
with the help of (\ref{6}) and (\ref{10}), using the primary definition
of Fox's H-function in \cite{37} with the integral identity [33, Eq. (3.194.3)], and then
utilizing [33, Eq. (8.384.1)], $I_{1}$ can be written as
\begin{align}
&I_{1}{=}\frac{\omega r\widetilde{\Xi}}{\overline\gamma_{1}\Gamma(m_w)\Gamma(k_w)}\frac{\gamma}{(2\pi i)^{2}}
{\int\limits_{\ell_{1}}}{\int\limits_{\ell_{2}}}\Gamma(t{-}s{-}1) && \nonumber\\
&{\times} \frac{\Gamma(1{+}rs)\Gamma({-}rs)\Gamma(1{+}s)}{\Gamma(1{-}rs)}
\frac{\Gamma(t{+}k_{w}{-}1)\Gamma(t{+}m_{w}{-}1)}{\Gamma(t)} &&\nonumber\\
&{\times} \left(\frac{C}{\lambda^r\mu_{r}}\right)^{-s}
\left(\frac{\widetilde{\Xi} \gamma}{\overline\gamma_{1}}\right)^{-t}dsdt&& \nonumber\\
&{+}\frac{(1{-}\omega) r\widetilde{\Xi}}{\overline\gamma_{1}\Gamma(m_w)\Gamma(k_w)\Gamma(a)c}
\frac{\gamma}{(2\pi i)^{2}}{\int\limits_{\ell_{1}}}
{\int\limits_{\ell_{2}}}\Gamma(t{-}s{-}1) &&\nonumber\\
&{\times} \frac{\Gamma\left(a{+}\frac{r}{c}s\right)\Gamma({-}\frac{r}{c}s)
\Gamma(1{+}s)}{\Gamma(1{-}\frac{r}{c}s)}
\frac{\Gamma(t{+}k_{w}{-}1)\Gamma(t{+}m_{w}{-}1)}{\Gamma(t)}&&\nonumber\\
&{\times} \left(\frac{C}{b^r\mu_{r}}\right)^{-s}
\left(\frac{\widetilde{\Xi} \gamma}{\overline\gamma_{1}}\right)^{-t}dsdt, &&
\tag{46}\label{46}
\end{align}
where $\ell_{1}$ and $\ell_{2}$ are two integral contours in
the complex domain. Then, based on the definition of the EGBFHF
[37, eq. (2.57)], the CDF can be derived as given in (\ref{13}).

\section{PDF of The End-to-End SNR for Fixed-Gain AF Relaying}
Based on the transformation in (\ref{14}) and using the same method as in \cite{43}, we have
\begin{align}
f_{\gamma^{AF}}(\gamma)
=&\frac{d}{d\gamma}\left[{\int_{0}^{\gamma}
\Pr\left[\gamma_{2}(\gamma_{1}-\gamma)<C\gamma\right]
f_{\gamma_{1}}(\gamma_{1})d\gamma_{1}}\right.\nonumber\\
&\left.{+\int_{\gamma}^{\infty}\Pr\left[\gamma_{2}(\gamma_{1}-\gamma)<C\gamma\right]
f_{\gamma_{1}}(\gamma_{1})d\gamma_{1}}\right],\tag{47}\label{47}
\end{align}
where $f_{\gamma_{1}}(\gamma_{1})$ is the PDF of $\gamma_{1}$. When $0<x<\gamma$,
$\Pr\left[\gamma_{2}(\gamma_{1}-\gamma)<C\gamma\right]=1$. Hence, (\ref{47}) can be written as
\begin{align}
f_{\gamma^{AF}}(\gamma)=&\frac{d}{d\gamma}\left[{\int_{0}^{\gamma}
f_{\gamma_{1}}(\gamma_{1})d\gamma_{1}}\right.\nonumber\\
&\left.{+\int_{\gamma}^{\infty}\Pr\left[\gamma_{2}(\gamma_{1}-\gamma)<C\gamma\right]
f_{\gamma_{1}}(\gamma_{1})d\gamma_{1}}\right].\tag{48}\label{48}
\end{align}
After calculating the derivative, we obtain
\begin{align}
f_{\gamma^{AF}}(\gamma)=f_{\gamma_{1}}(\gamma)-\lim_{\gamma_{1}\rightarrow \gamma^{+}}\Pr\left[\gamma_{2}<\frac{C\gamma}{(\gamma_{1}-\gamma)}\right]f_{\gamma_{1}}(\gamma)\nonumber\\
+\int_{\gamma}^{\infty}
\frac{C\gamma_{1}}{(\gamma_{1}-\gamma)^2}
f_{\gamma_{2}}\left(\frac{C\gamma}{\gamma_{1}-\gamma}\right)
f_{\gamma_{1}}(\gamma_{1})d\gamma_{1}.
\tag{49}\label{49}
\end{align}
Since $\lim_{\gamma_{1}\rightarrow \gamma^{+}}\Pr[\gamma_{2}<C\gamma/(\gamma_{1}-\gamma)]=
\Pr[\gamma_{2}<+\infty]=1$, we have
\begin{align}
f_{\gamma^{AF}}(\gamma)=\int_{\gamma}^{\infty}
\frac{C\gamma_{1}}{(\gamma_{1}-\gamma)^2}
f_{\gamma_{2}}\left(\frac{C\gamma}{\gamma_{1}-\gamma}\right)
f_{\gamma_{1}}(\gamma_{1})d\gamma_{1}.\tag{50}\label{50}
\end{align}
By using (\ref{6}) and (\ref{9}), (\ref{50}) simplifies to
\begin{align}
&f_{\gamma^{AF}}(\gamma){=}\frac{\omega \widetilde{\Xi}}
{\overline\gamma_{1}\Gamma(m_w)\Gamma(k_w)}
\frac{1}{(2\pi i)^{2}}\int\limits_{\ell_{1}}\int\limits_{\ell_{2}}
\Gamma(t{-}s{-}1)\nonumber\\
&\times \Gamma(1{+}rs)\Gamma(s) \frac{\Gamma(t{+}k_{w}{-}1)\Gamma(t{+}m_{w}{-}1)}{\Gamma(t{-}1)}\nonumber\\
&\times\left(\frac{C}{\lambda^r\mu_{r}}\right)^{-s}
\left(\frac{\widetilde{\Xi} \gamma}{\overline\gamma_{1}}\right)^{-t}dsdt \nonumber\\
&+\frac{(1-\omega)\widetilde{\Xi}}{\overline\gamma_{1}
\Gamma(m_w)\Gamma(k_w)\Gamma(a)}\frac{1}{(2\pi i)^{2}}
\int\limits_{\ell_{1}}\int\limits_{\ell_{2}}\Gamma(t{-}s{-}1)\nonumber\\
&\times \Gamma\left(a+\frac{r}{c}s\right)\Gamma(s)
\frac{\Gamma(t{+}k_{w}{-}1)\Gamma(t{+}m_{w}{-}1)}{\Gamma(t{-}1)}\nonumber\\
&\times \left(\frac{C}{b^r\mu_{r}}\right)^{-s}
\left(\frac{\widetilde{\Xi} \gamma}{\overline\gamma_{1}}\right)^{-t}dsdt.
\tag{51}\label{51}
\end{align}
Using the same method as in Appendix A, (\ref{15}) is obtained.

\section{Asymptotic Outage for Fixed-Gain AF Relaying}
By using the definition of the EGBFHF in [45, Eq. (1.1)], $I_{1}$ can be written as
\begin{align}
&I_{1}{=}\frac{\omega r\widetilde{\Xi}\gamma}{\overline\gamma_{1}\Gamma(m_w)\Gamma(k_w)}\frac{1}{(2\pi i)^{2}}
{\int\limits_{\ell_{1}}}{\int\limits_{\ell_{2}}} && \nonumber\\
&{\times} \, {\mathrm{H}}_{1,3}^{3,0}
\left [{{\frac{\widetilde{\Xi}\gamma}{\overline\gamma_{1}}}\left |{ \begin{matrix} {(0,1)}
\\ {({-}s{-}1,1)(k_{w}{-}1,1)(m_{w}{-}1,1)} \\ \end{matrix} }\right . }\right ]\!\nonumber\\
&\times \frac{\Gamma(1{+}rs)\Gamma({-}rs)\Gamma(1{+}s)}{\Gamma(1{-}rs)}\left(\frac{C}{\lambda^r\mu_{r}}\right)^{-s} ds &&\nonumber\\
&{+}\frac{(1{-}\omega) r\widetilde{\Xi}\gamma}{\overline\gamma_{1}\Gamma(m_w)\Gamma(k_w)\Gamma(a)c}\frac{1}{(2\pi i)^{2}}
{\int\limits_{\ell_{1}}}{\int\limits_{\ell_{2}}} && \nonumber\\
&{\times} \, {\mathrm{H}}_{1,3}^{3,0}
\left [{{\frac{\widetilde{\Xi}\gamma}{\overline\gamma_{1}}}\left |{ \begin{matrix} {(0,1)}
\\ {({-}s{-}1,1)(k_{w}{-}1,1)(m_{w}{-}1,1)} \\ \end{matrix} }\right . }\right ]\!\nonumber\\
&\times \frac{\Gamma\left(a{+}\frac{r}{c}s\right)\Gamma\left({-}\frac{r}{c}s\right)
\Gamma(1{+}s)}{\Gamma\left(1{-}\frac{r}{c}s\right)}\left(\frac{C}{b^r\mu_{r}}\right)^{-s} ds &&\nonumber\\
&{=}\Lambda_{1}+\Lambda_{2}.
\tag{52}\label{52}
\end{align}
By letting $\overline{\gamma}_{1}\rightarrow \infty$, the Fox's H-functions in (\ref{52}) can
be approximated by using its Taylor expansion in [46, Eq. (1.8.4)]
as
\begin{align}
\, {\mathrm{H}}_{1,3}^{3,0}
\left [{{\frac{\widetilde{\Xi}\gamma}{\overline\gamma_{1}}}\left |{ \begin{matrix} {(0,1)}
\\ {({-}s{-}1,1)(k_{w}{-}1,1)(m_{w}{-}1,1)} \\ \end{matrix} }\right . }\right ]\!\nonumber\\
\approx \frac{\Gamma(k_{w}{+}s)\Gamma(m_{w}{+}s)}{\Gamma(1{+}s)}
\left(\frac{\widetilde{\Xi}\gamma}{\overline\gamma_{1}}\right)^{({-}s{-}1)}.\tag{53}\label{53}
\end{align}
By substituting (\ref{53}) into (\ref{52}) and applying [45, Eq. (1.1)], $\Lambda_{1}$ and $\Lambda_{2}$ can be rewritten as
\begin{align}
\Lambda_{1} {=} \frac{wr}{\Gamma(m_{w})\Gamma(k_{w})}\, {\mathrm{H}}_{1,4}^{3,1}
\left [{{\frac{C\widetilde{\Xi}\gamma}{\overline\gamma_{1}\lambda^r\mu_r}}\left |{ \begin{matrix} {(1,r)}
\\ {(1,r)(k_{w},1)(m_{w},1)(0,r)} \\ \end{matrix} }\right . }\right ]\!,\tag{54}\label{54}
\end{align}
and
\begin{align}
\Lambda_{2} &{=} \frac{(1{-}w)r}{\Gamma(m_{w})\Gamma(k_{w})\Gamma(a)c}\nonumber\\
&\times \, {\mathrm{H}}_{1,4}^{3,1}
\left [{{\frac{C\widetilde{\Xi}\gamma}{\overline\gamma_{1}b^r\mu_r}}\left |{ \begin{matrix} {(1,\frac{r}{c})}
\\ {(a,\frac{r}{c})(k_{w},1)(m_{w},1)(0,\frac{r}{c})} \\ \end{matrix} }\right . }\right ]\!,\tag{55}\label{55}
\end{align}
By letting $\mu_{r}\rightarrow \infty$ and utilizing [46, Eq. (1.8.4)], $\Lambda_{1}$ can be further simplified as
\begin{align}
&\Lambda_{1}\approx \frac{w\Gamma\left(k_w{-}\frac{1}{r}\right)\Gamma\left(m_w{-}\frac{1}{r}\right)}
{\Gamma(m_{w})\Gamma(k_{w})}\left(\frac{C\widetilde{\Xi} \gamma}{\lambda^r\mu_{r}
\overline{\gamma}_{1}}\right)^{\frac{1}{r}}\nonumber\\
&+\frac{wr\Gamma\left(1{-}k_{w}r\right)\Gamma\left(m_w{-}k_w\right)\Gamma(k_{w}r)}
{\Gamma(m_{w})\Gamma(k_{w})\Gamma(1{+}k_{w}r)}\left(\frac{C\widetilde{\Xi}\gamma}{\lambda^r\mu_{r}
\overline{\gamma}_{1}}\right)^{k_w}\nonumber\\
&+\frac{wr\Gamma\left(1{-}m_{w}r\right)
\Gamma\left(k_w{-}m_w\right)\Gamma(m_{w}r)}{\Gamma(m_{w})\Gamma(k_{w})\Gamma(1{+}m_{w}r)}
\left(\frac{C\widetilde{\Xi}\gamma}{\lambda^r\mu_{r}\overline{\gamma}_{1}}\right)^{m_w}.\tag{56}\label{56}
\end{align}
Following the same approach as for $\Lambda_{1}$ and
applying [45, Eq. (1.1)] and [46, Eq. (1.8.4)] with the aid of some algebraic manipulations,
we get the asymptotic expression of $\Lambda_{2}$. With the help of (\ref{21}) and the
asymptotic expressions of $\Lambda_{1}$ and $\Lambda_{2}$, we obtain the closed-form expression for the OP at high SNR in (\ref{22}).

\section{ABER for Fixed-Gain AF Relaying}
Based on (\ref{46}) and utilizing [33, Eq. (3.326.2)], we have
\begin{align}
&I_{2}=\frac{\omega r\widetilde{\Xi}}{2\overline\gamma_{1}\Gamma(p)
\Gamma(m_w)\Gamma(k_w)q}\frac{1}{(2\pi i)^{2}}\int\limits_{\ell_{1}}
\int\limits_{\ell_{2}}\Gamma(t{-}s{-}1)\nonumber\\
&\times \frac{\Gamma(1{+}rs)\Gamma({-}rs)\Gamma(1{+}s)}{\Gamma(1{-}rs)}\left(\frac{C}{\lambda^r\mu_{r}}\right)^{-s}\nonumber\\
&\times \frac{\Gamma(t{+}k_{w}{-}1)\Gamma(t{+}m_{w}{-}1)\Gamma(p{+}1{-}t)}{\Gamma(t)}
\left(\frac{\widetilde{\Xi}}{\overline\gamma_{1}q}\right)^{-t} dsdt\nonumber\\
&+\frac{(1-\omega)r\widetilde{\Xi}}{2\overline\gamma_{1}\Gamma(p)
\Gamma(m_w)\Gamma(k_w)\Gamma(a)cq}\frac{1}{(2\pi i)^{2}}
\int\limits_{\ell_{1}}\int\limits_{\ell_{2}}\Gamma(t{-}s{-}1)\nonumber\\
&\times \frac{\Gamma\left(a+\frac{r}{c}s\right)\Gamma\left(-\frac{r}{c}s\right)
\Gamma(1{+}s)}{\Gamma\left(1-\frac{r}{c}s\right)}\left(\frac{C}{b^r\mu_{r}}\right)^{-s}\nonumber\\
&\times \frac{\Gamma(t{+}k_{w}{-}1)\Gamma(t{+}m_{w}{-}1)}{\Gamma(t)}
\left(\frac{\widetilde{\Xi}}{\overline\gamma_{1}q}\right)^{-t} dsdt.\tag{57}\label{57}
\end{align}
By using [37, Eq. (2.57)], $I_{2}$ is obtained as in (\ref{27}). Finally, with the help of (\ref{26}) and (\ref{27}), the ABER can be formulated in closed-form in terms of the EGBHFH.

\end{document}